\documentclass[prl,twocolumn,showpacs,preprintnumbers,amsmath,amssymb,superscriptaddress]{revtex4}

\usepackage{color}
\usepackage{graphicx}
\usepackage{dcolumn}
\usepackage{bm}
\usepackage{hyperref}

\begin{document}


\title{Correlation based entanglement criteria for bipartite systems}

\author{Yumang Jing}
\email{jingyumang@pku.edu.cn}
\affiliation{State Key Laboratory of Mesoscopic Physics, School of Physics, Peking University, Collaborative Innovation Center of Quantum Matter, Beijing 100871, China}

\author{Qiongyi He}
\affiliation{State Key Laboratory of Mesoscopic Physics, School of Physics, Peking University, Collaborative Innovation Center of Quantum Matter, Beijing 100871, China}
\affiliation{Collaborative Innovation Center of Extreme Optics, Shanxi University, Taiyuan, Shanxi 030006, China}

\author{Tim Byrnes}
\affiliation{New York University Shanghai, 1555 Century Ave, Pudong, Shanghai 200122, China}  
\affiliation{NYU-ECNU Institute of Physics at NYU Shanghai, 3663 Zhongshan Road North, Shanghai 200062, China}
\affiliation{National Institute of Informatics, 2-1-2 Hitotsubashi, Chiyoda-ku, Tokyo 101-8430, Japan}
\affiliation{Department of Physics, New York University, New York, NY 10003, USA}

\date{\today}

\begin{abstract}
We introduce a class of inequalities based on low order correlations of operators to detect entanglement in bipartite systems. The operators may either be Hermitian or non-Hermitian and are applicable to any physical system or class of states. Testing the criteria on example systems reveals that they outperform other common correlation based criteria, such as those by Duan-Giedke-Cirac-Zoller and Hillery-Zubairy.  One unusual feature of the criteria is that the correlations include the density matrix itself, which is related to the purity of the system.  We discuss how such a term could be measured in relation to the criteria.  
\end{abstract}

\pacs{03.75.Dg, 37.25.+k, 03.75.Mn}
            
\maketitle

The generation of entanglement is an essential task in quantum information science, and is fundamental to any classically intractable task such as quantum teleportation or quantum computation. Detecting entanglement is therefore an important task that must necessarily be carried out in this context, and is required for benchmarking and characterizing the quantum states created.  
The simplest system for studying entanglement is the bipartite system. The Peres-Horodecki criterion was first proposed as a necessary condition for all bipartite separable states, and sufficient as well in $2\times2$ or $2\times3$ dimensional systems \cite{peres1996, horodecki1996, horodecki1997}. Beyond such low dimensional systems, sufficient inseparability conditions based on second-order moments have been derived \cite{duan,simon}, which have also been shown necessary for the special case of Gaussian states. However, non-Gaussian states are also crucial in some cases. Some of the entanglement criteria derived so far are based on some forms of uncertainty relations \cite{hofmann2003,o.guhne,toth,hz2006,Eric11,PQS11,BEC11}, and especially, in certain cases, in conjunction with partial transposition \cite{vogel2005} and via SU(1,1) and SU(2) algebra as well \cite{ab-newj2005,nha-kim,nhaSR} for non-Gaussian states.  

What is particularly useful in the context of experimental verification of entanglement are correlation based witnesses, where a small number of observables are measured, and entanglement can be verified.  Examples of such correlation based methods include those by Duan-Giedke-Cirac-Zoller (DGCZ) \cite{duan} and Hillery-Zubairy \cite{hz2006}.  These are used when full tomography of the density matrix is impractical or impossible, and thus only incomplete information of the system is available. Such correlation based methods are typically only a sufficient condition for entanglement, and can fail to detect entanglement across a broad class of entangled states.  In this paper, we will introduce another class of correlation based entanglement criteria, which work without assumption of the class of states (Gaussian or non-Gaussian).  It is applicable to any physical system and is defined in terms of low-order correlations of Hermitian or non-Hermitian operators. These inequalities do not use any special properties of annihilation operators, allowing for a rather general purpose entanglement witness.  Testing the criterion for some typical states we find that it works in a wider range of parameters than similar correlation based methods such as those by DGHZ and Hillery-Zubairy.  The use of any type of operator in the criterion is particularly advantageous in comparison to these approaches where certain assumptions need to be satisfied by the observables.  The cost of this improvement is that the criteria involves an average over the density matrix squared which can be related to the purity of the system.  We discuss how the evaluation of such a term can be evaluated in an experimental setting, together with the performance.


We first introduce and prove the entanglement criterion. Consider a system consisting of two subsystems which is described by a Hilbert space $\mathcal{H}=\mathcal{H}_{a}\otimes\mathcal{H}_{b}$. Let $A_{i}$ be any operator on $\mathcal{H}_{a}$ and $B_{i}$ be any operator on $\mathcal{H}_{b}$ with $i \in \{ 1,2 \}$.  Consider the density matrix of a general separable state in diagonal form
$\rho=\sum_{k}p_{k} |k \rangle_a \langle k |_a  \otimes |k \rangle_b \langle k|_b  $, where $p_{k}$ is a probability $\sum_{k}p_{k}=1$. The states $ |k \rangle_{a,b}  $ on subsystems $a $ or $ b $ are not necessarily orthogonal $ \langle k |k' \rangle_{a,b} \ne \delta_{k k'}$, but are normalized.  We define
\begin{align}
U= (A_1)^{\sigma_{A_1}}  |k \rangle_a \langle k |_a    (A_1^{\dagger})^{\bar{\sigma}_{A_1}} \otimes (B_1)^{\sigma_{B_1}}  |l \rangle_b \langle l|_b   (B_{1}^{\dagger})^{\bar{\sigma}_{B_1}}  , \nonumber\\
V=(A_2)^{\sigma_{A_2}} |l \rangle_a \langle l |_a   (A_{2}^{\dagger})^{\bar{\sigma}_{A_2}} \otimes (B_2)^{\sigma_{B_2}} |k \rangle_b \langle k|_b  (B_{2}^{\dagger} )^{\bar{\sigma}_{B_2}} .
\label{def}
\end{align}
The $ \sigma_{A_i,B_i} =\{ 0,1 \} $ are binary parameters which serve to adjust the position of the operators and $ \bar{\sigma}_{A_i,B_i}  = 1- \sigma_{A_i,B_i}  $.  Now consider the quantity $Z^{\dagger}Z$, where $Z= U- e^{i\phi} V$ and $ \phi $ is a free parameter.  Since $Z^{\dagger}Z$ is a semi-positive Hermitian operator, which means $\text{Tr}(Z^{\dagger}Z) \geq 0$, and $\sum_{k,l}p_k p_l \text{Tr}(Z^{\dagger}Z) \geq 0$ will always hold. Then for separable states, we have
\begin{align}
&\langle n_{A_{1}} \rangle\langle n_{B_{1}}  \rangle+\langle n_{A_{2}} \rangle\langle n_{B_{2}} \rangle \nonumber\\
&-e^{i \phi}  \text{Tr} \Big[ (A_{1}^{\dagger} )^{\sigma_{A_1}}  (A_{2})^{\sigma_{A_2}}  (B_{2}^{\dagger})^{\bar{\sigma}_{B_2}}  (B_{1})^{\bar{\sigma}_{B_1}} \rho  \nonumber\\
& \times (A_{2}^{\dagger})^{\bar{\sigma}_{A_2}}  (A_{1})^{\bar{\sigma}_{A_1}}   (B_{1}^{\dagger})^{\sigma_{B_1}}  (B_{2})^{\sigma_{B_2}}  \rho \Big] \nonumber\\
&- e^{-i \phi} \text{Tr} \Big[ (A_{2}^{\dagger})^{\sigma_{A_2}}  ( A_{1})^{\sigma_{A_1}}  (B_{1}^{\dagger})^{\bar{\sigma}_{B_1}}   (B_{2})^{\bar{\sigma}_{B_2}} \rho  \nonumber\\
& (A_{1}^{\dagger})^{\bar{\sigma}_{A_1}}  (A_{2})^{\bar{\sigma}_{A_2}} (B_{2}^{\dagger})^{\sigma_{B_2}}  (B_{1})^{\sigma_{B_1}}  \rho \Big] \geq 0.
\label{proof}
\end{align}
where $n_{A_{i}}=A_{i}^{\dagger}A_{i}$ and $n_{B_{i}}=B_{i}^{\dagger}B_{i}$. As $A_{i}$ and $B_{i}$ are arbitrary operators, which can be either Hermitian or non-Hermitian, $ n$ may or may not be number operators. Eq. (\ref{proof}) is true for any separable state.  Hence, any violation of (\ref{proof}) shows that a state is entangled.

Similarly, we may define
\begin{align}
U& = (A_1)^{\sigma_{A_1}}   |k \rangle_a \langle k |_a  (A_1^{\dagger})^{\bar{\sigma}_{A_1}} \otimes (B_1)^{\sigma_{B_1}}  |k \rangle_b \langle k|_b    (B_{1}^{\dagger})^{\bar{\sigma}_{B_1}}  , \nonumber\\
V& =(A_2)^{\sigma_{A_2}}|l \rangle_a \langle l |_a   (A_{2}^{\dagger})^{\bar{\sigma}_{A_2}} \otimes (B_2)^{\sigma_{B_2}} |l \rangle_b \langle l|_b    (B_{2}^{\dagger} )^{\bar{\sigma}_{B_2}} .
\label{def2}
\end{align}
which is the same as (\ref{def}) except that the $ k$ and $ l$ labels are interchanged on subsystem $ b $.  Using the same steps we obtain the inequality
\begin{align}
&\langle n_{A_{1}} n_{B_{1}}  \rangle+\langle n_{A_{2}}  n_{B_{2}} \rangle \nonumber\\
&- e^{i \phi}  \text{Tr} \Big[ (A_{1}^{\dagger} )^{\sigma_{A_1}}  (A_{2})^{\sigma_{A_2}}  (B_{1}^{\dagger})^{\sigma_{B_1}}  (B_{2})^{\sigma_{B_2}}  \rho  \nonumber\\
& \times (A_{2}^{\dagger})^{\bar{\sigma}_{A_2}}  (A_{1})^{\bar{\sigma}_{A_1}}  (B_{2}^{\dagger})^{\bar{\sigma}_{B_2}}  (B_{1})^{\bar{\sigma}_{B_1}} \rho   \Big] \nonumber\\
&- e^{-i \phi}  \text{Tr} \Big[ (A_{2}^{\dagger})^{\sigma_{A_2}}  ( A_{1})^{\sigma_{A_1}}   (B_{2}^{\dagger})^{\sigma_{B_2}}  (B_{1})^{\sigma_{B_1}}  \rho \nonumber\\
& (A_{1}^{\dagger})^{\bar{\sigma}_{A_1}}  (A_{2})^{\bar{\sigma}_{A_2}}  (B_{1}^{\dagger})^{\bar{\sigma}_{B_1}}   (B_{2})^{\bar{\sigma}_{B_2}} \rho \Big] \geq 0.
\label{proof2}
\end{align}
Again, any violation of (\ref{proof2}) shows that the state is entangled.  

The inequalities (\ref{proof}) and (\ref{proof2}) are our main result.  To illustrate their utility, let us consider a few special cases.  Setting $A_i=A$, $B_{i}=B$, $ \sigma_{A_i} = 1 $, $ \sigma_{B_i} = 0 $, and $ \phi = 0 $, Eq. (\ref{proof}) reduces to
\begin{align}
\langle n_{A}\rangle\langle n_{B}\rangle \geq \langle n_{A} n_{B}\rho\rangle.
\label{finalcri}
\end{align}
where $ \langle X \rho Y \rangle \equiv \text{Tr} ( X \rho Y \rho) $ for any operators $ X,Y $. Another example is $A_i=A$, $B_{i}=B$, and  $ \sigma_{A_1} = \sigma_{B_2} =  1 $, $\sigma_{A_2} = \sigma_{B_1} = 0 $ in Eq. (\ref{proof2}), for which we obtain
\begin{align}
2 \langle n_{A} n_{B}\rangle \geq e^{i \phi} \langle A^\dagger B \rho A^\dagger B \rangle +  e^{-i \phi} \langle A B^\dagger \rho A B^\dagger \rangle .
\label{finalcri2}
\end{align}
We can thus generate a whole family of inequalities which all serve as entanglement witnesses by various choices of $ \sigma $ and operators.  The phase $ \phi $ can be chosen in such a way such that the last two terms in (\ref{proof}) and (\ref{proof2}) take its largest negative value, which gives the best chance of violating the inequality.  We note that for any pure state, we have $\rho^2=\rho$ so that (\ref{finalcri}) reduces to the ordinary correlation function $ \langle n_{A}\rangle\langle n_{B}\rangle-\langle n_A n_B\rangle$, which has known connections to entanglement for pure states \cite{barnum, somma}. Meanwhile (\ref{finalcri2}) reduces to $ 2 \langle n_{A} n_{B}\rangle \geq e^{i \phi} \langle A^\dagger B \rangle^2  +  e^{-i \phi} \langle A B^\dagger \rangle^2 $.


The unusual feature of the criteria that we consider in this paper is that the density matrix is involved in the correlations themselves.  The origin of this is that in the terms involving a trace in (\ref{proof}) and (\ref{proof2}) there are two density matrices. While this makes evaluation of the correlations slightly less convenient than an ordinary expectation value, this helps to make our entanglement criterion more sensitive than existing correlation based criteria.  On first glance it may appear that evaluation of such a term would require knowledge of the full density matrix via tomography, which defeats the purpose of a correlation based entanglement witness.  We now show two strategies that can estimate this term without full tomography.  We will focus upon the specific case of (\ref{finalcri}) as this is the criterion which we have found to be simplest and also most sensitive for various states that we have examined.  

The first of the methods involves a measurement of the observable $ X = n_{A} n_{B} $ combined with some simple post-processing.  Our aim will be to obtain an estimator for $\langle X \rho\rangle $, which we will call $ {\cal E} (X,\rho) $.  The aim of any estimator is that it should give a reasonable approximation to the desired quantity, i.e. $ {\cal E} (X,\rho) \approx \langle X \rho\rangle $. Furthermore, in the context of the 
criterion (\ref{finalcri}) we would like that $ {\cal E} (X,\rho) \le \langle X \rho\rangle $, in order that estimator can replace the $\langle n_{A} n_{B}\rho\rangle $ term.  Let $X$ be an operator which can be expanded in terms of its eigenstates as $X=\sum_n X_{n}| X_n\rangle \langle X_n | $, where $| X_n\rangle$ and $ X_n $ are the $ n $th eigenstate and eigenvalue respectively.  Now consider making a measurement of the state $ \rho $ with respect to $ X $.  The particular outcome $ n $ will be obtained with probability $P_{n}=\text{Tr} ( | X_n\rangle \langle X_n | \rho)$. We then propose that an estimator for $\langle X \rho\rangle $ is
\begin{align}
{\cal E} (X,\rho) = \sum_n P_{n}^2 X_{n}. 
\label{estimator}
\end{align}
To show that the estimator has the desired properties, write the density matrix in its diagonal form  $\rho=\sum_k p_{k}| \Psi_{k}\rangle\langle\Psi_{k}| $.  We may evaluate that
\begin{align}
\langle X \rho\rangle  & = \sum_{n} \sum_{k} p_{k}^{2} | \langle \Psi_{k} |  X_n \rangle|^{2} X_{n}, \label{real} \\
{\cal E} (X,\rho) & =\sum_{n} \big(\sum_{k}  p_{k} | \langle \Psi_{k} |  X_n \rangle |^{2} \big)^{2} X_{n}  .
\end{align}
Under the condition that $ |  X_n \rangle = | \Psi_{n}\rangle $, these two expressions coincide.  Such conditions can be satisfied by either making a choice of measurement $ A, B $ such that the eigenstates of $ n_A n_B $ coincides with the state being prepared.  Alternatively, for very mixed states $\rho \approx I/d$ where $ d $ is the dimension of the system, the choice of basis can be made arbitrarily, and the estimator agrees with the desired expression.  Thus we expect such an estimator to give a good approximation for states with low purity.  By looking at the difference $ \langle X \rho\rangle  - {\cal E} (X,\rho) $ and taking $ | \langle \Psi_{k} |  X_n \rangle|^{2} $ to be a probability distribution, it can be shown that $  {\cal E} (X,\rho) \le \langle X \rho\rangle $ as it is a sum of variances. Thus (\ref{estimator}) has the desired properties of an estimator and does not require a full tomography of the density matrix. 

The second approach is based on the observation that  $\langle X \rho\rangle = \text{Tr} (X \rho^2 ) $ is a quantity that is closely related to the purity $ \text{Tr} (\rho^2) $.  We may then construct another estimator
\begin{align}
{\cal E} (X,\rho) = \langle X \rangle \text{Tr} (\rho^{2}),
\label{estimator2}
\end{align}
which is a mean field approximation in $ X $ and $ \rho $.  Since for a pure state $ \langle X \rho\rangle =\langle X \rangle $, we expect that this estimator should work in the opposite limit to that of (\ref{estimator}), which is more appropriate for strongly mixed states. Again in the context of our criterion we require that the estimator 
underestimate the actual  $ \langle X \rho\rangle $.  While this is not generally true of (\ref{estimator2}), under particular conditions this may be satisfied.  To see this, consider a nearly pure state
\begin{align}
\rho=p|\Psi_{0}\rangle\langle\Psi_{0}|+(1-p)|\Psi_{1}\rangle\langle\Psi_{1}|,
\end{align}
where $ |\Psi_{0}\rangle $ can be thought as being the target state, and $ |\Psi_{1}\rangle $ is some undesired noise state. For this case we can evaluate that
\begin{align}
\langle X \rho\rangle -{\cal E} (X,\rho) =(1-p)p(2p-1)( \langle X \rangle_0 - \langle X \rangle_1 ),
\end{align}
where $ \langle X \rangle_k = \langle  \Psi_{k} | X | \Psi_{k} \rangle $.  For $ 1/2 < p \le 1 $, this is a positive quantity if $ \langle X \rangle_{0}\geq\langle X\rangle_{1} $ as desired.  In this case we can use the estimator (\ref{estimator2}) conditionally, when some information of the desired and noise states of the system are known.  

Writing the estimator in the form of (\ref{estimator2}) requires an estimate of the purity of the system $ \text{Tr} (\rho^{2}) $.  This has been investigated in many past works, we quote several approaches here which give a relatively simple way of estimating it. The purity may be obtained by summing over variances or expectation values of observables according to
\begin{align}
\text{Tr} (\rho^2) & = d - \sum_l \delta (M_l)^2 = \sum_l  |\langle M_l \rangle|^2
\end{align}
where the observables $ M_k $ satisfy $\text{Tr} (M_{k} M_{l})=\delta_{kl} $ \cite{Guhne}. This is tractable for low-dimensional systems such as qubits, but for infinite dimensional systems such as photons is unsuitable due to the large number of observables.  For such systems, an estimate of the purity may be obtained from the covariance matrix $ C $.  For photonic Gaussian states the purity is given by $ \text{Tr} (\rho^2) = 1/\sqrt{\text{Det}(C)} $ where for a two mode state $ C $ is the submatrix with cross-correlations between the modes \cite{tahira09,kogias15,Manko}. In addition, several theoretical and experimental works show methods for directly measuring purity by creating two copies of the system, and interfering them with each other \cite{nature,Karol}.

\begin{figure}
\includegraphics[width=\columnwidth]{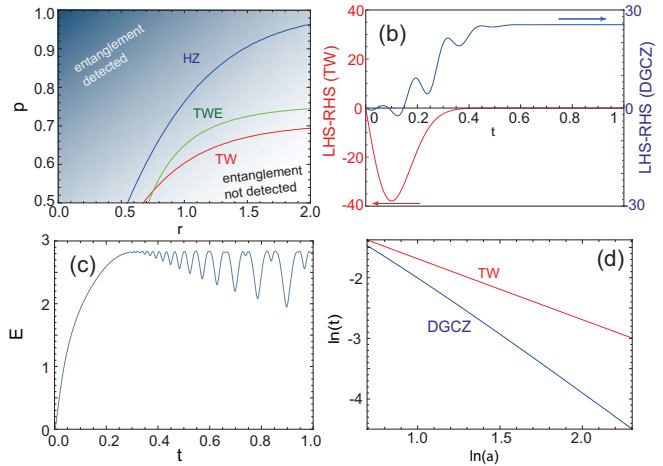}
\caption{\label{fig1} (a) Regions of parameter space for the mixed two mode squeezed state (\ref{thermaltwomode}) where entanglement is detected for several criteria as marked.  HZ corresponds to the Hillery-Zubairy criterion $ | \langle ab\rangle |^2> \langle n_{a}\rangle\langle n_{b}\rangle$ (entangled); TW (this work) corresponds to (\ref{finalcri}); TWE (this work with estimator) corresponds to using (\ref{finalcri}) with (\ref{estimator2}).  (b) Violation of the entanglement criteria for evolution of two coherent states $ | \alpha \rangle_a | \beta \rangle_b $ under a cross-Kerr Hamiltonian $ H = n_a n_b $ for a time $ t $.  For the two criteria (\ref{finalcri})  and (\ref{duankerr}) the LHS-RHS is plotted, such that a violation of the inequality (and therefore entanglement) is negative.  Parameters used are $ \alpha = \beta = 5 $. (c) Entropy of entanglement as a function of time $t$ for cross-Kerr Hamiltonian, with $\alpha=\beta=5$. (d) Scaling of the criteria (\ref{finalcri}) and (\ref{duankerr}) as a function of $ \alpha = \beta $.  TW corresponds to the time where LHS-RHS is a minimum for  (\ref{finalcri}); DGCZ corresponds to the time where (\ref{duankerr}) has LHS=RHS. }
\end{figure}

We now apply our criterion to several examples and compare the performance with other correlation based entanglement witnesses.  
In our first example we apply our methods to the Bell states
\begin{align}
|{\cal B}_1 \rangle =\frac{1}{\sqrt{2}}(| 1 \rangle_{a} | 0 \rangle_{b}+ | 0 \rangle_{a}| 1 \rangle_{b}), \nonumber \\
|{\cal B}_2 \rangle =\frac{1}{\sqrt{2}}(| 0 \rangle_{a} | 0 \rangle_{b}+ | 1 \rangle_{a}| 1 \rangle_{b}). 
\end{align}
Specifically, we apply the criterion (\ref{finalcri}) taking  $A = \sigma_{a}^{-} $ and $B = \sigma_{b}^{-} $, where $ \sigma^{-} $ is a Pauli spin lowering operator.  For either Bell state, simple evaluation of (\ref{finalcri}) shows a violation of the inequality, signalling entanglement.  We note that using the Hillery-Zubairy criteria, while $ |{\cal B}_1 \rangle $ shows entanglement, $ |{\cal B}_2 \rangle $ does not, as these are treated using two separate inequalities.  In an optics setting, $ |{\cal B}_1 \rangle $  corresponds to beam splitter type of entanglement, while $ |{\cal B}_2 \rangle $ corresponds to parametric down conversion, which have two different criteria for the Hillery-Zubairy approach. 

For mixed qubit states, we consider the Werner state $\rho_i= p|{\cal B}_i \rangle\langle{\cal B}_i |+ I (1-p)/4 $, where $0\leq p\leq1$, $ i = 1,2 $, and $I$ is the identity operator. Using our criteria (\ref{finalcri}) we obtain that the state is entangled in the range $ 0.6 <p \leq 1$.  For comparison, the Hillery-Zubairy criterion gives entanglement in the range $(\sqrt{5}-1)/2 \hspace{2mm}  (\approx 0.62) < p \leq1$ (the criterion (\ref{finalcri2}) gives exactly this same range). Thus our criterion (\ref{finalcri}) provides a slightly wider range of entanglement detection for qubits.  We note that other criteria such the Peres-Horodecki (positivity under partial transpose) criterion \cite{horodecki1996} do give a larger range of entanglement detection $ 1/3 < p \le 1 $. However, this is not a correlation based method and can be expected in general to perform better as it uses the complete information available in the density matrix. 

The second example is a mixed state consisting of a two-mode squeezed vacuum state $ | r \rangle $ and a thermal state on each mode $ \rho_{\text{th}}^{a,b} $, with density matrix 
\begin{align}
\rho& =p |r\rangle \langle r| +(1-p) \rho_{\text{th}}^a \otimes \rho_{\text{th}}^b  \label{thermaltwomode} \\
|r\rangle & =\sqrt{1-\tanh^{2}(r)} \sum_n \tanh^{n}(r) |n \rangle_a | n\rangle_b \nonumber \\
\rho_{\text{th}}^{a,b} & = (1-\tanh^{2}(r)) \sum_n \tanh^{2n}(r) |n \rangle_{a,b} \langle n |_{a,b} ,
\end{align}
where $r$ is the squeezing parameter, and we have assumed that the thermal state has the same thermal characteristics as the squeezing parameter due to some decoherence in the system.  The result of using our criterion (\ref{finalcri}) is shown in Fig. \ref{fig1}(a). From the calculation of the negativity \cite{vidal02,zyczkowski98}, we find that the state (\ref{thermaltwomode}) is always entangled for any $ p>0 $.  Our criterion shows that the state is entangled in a large portion of the parameter space, and indicates that it is a sensitive detector of entanglement. In comparison, the Hillery-Zubairy criterion shows a smaller range of parameters that reveal entanglement.  The dashed line in Fig. \ref{fig1}(a) represents the bound that would be obtained by the application of the second estimator (\ref{estimator2}).  While the range of entanglement is reduced, it still shows a larger range than the Hillery-Zubairy approach. We note the estimator must be used conditionally where the $ \langle n_A n_B \rangle $ is larger for the squeezed state than the thermal state and $ p> 1/2 $, hence works only in a restricted region.

The third example we show is entanglement due to a cross-Kerr nonlinearity \cite{schmidt96}, which exhibits non-Gaussian characteristics. Evolving the Hamiltonian $ H=n_a n_b $ for a time $ t $ gives Heisenberg equations as
\begin{align}
p_a(t) \approx p_a(0) - t x_a(0) n_b(0), \nonumber \\
x_a (t) \approx x_a(0) + t p_a(0) n_b(0), 
\end{align}
and similarly with $ a \leftrightarrow b $.  Taking two initially coherent states $| \alpha\rangle_a |\beta\rangle_b$ where $ \alpha, \beta $ are real and positive, we expect that initially correlations between $ p_a $ and $ n_b $ should develop, as $ x_a(0) \approx \alpha $ and $ x_b(0) \approx \beta $.  At later times when $ p_a $ is not necessarily small, we also expect that correlations between $ x_a $ and $ n_b $ should also be present.  More generally, we expect that correlations between operators
\begin{align}
A=\frac{1}{2}(e^{-i\theta} a^{\dagger}+ e^{i\theta} a ), \quad B=b \label{x1b2}
\end{align}
are present, where $ a,b $ are annihilation operators for mode $a$ and $b$, and $\theta$ is a free parameter that may be optimized.  $ B $ could equally be taken as $ n_b $, however, we find that the above choice works equally well (Fig. \ref{fig1}(b)). Our criterion detects entanglement for all $ t $ except $ t = 2\pi n $ where $ n $ is an integer, where the state becomes disentangled again.  This can be compared to a calculation of the entropy in Fig.\ref{fig1}(c) which reveals that entanglement is present for the same times.  The entropy has a fractal form that is reminiscent of the ''Devil's crevasse'' entanglement as seen in entangled spinor Bose-Einstein condensates \cite{byrnes13}. We note that the violation is only significant for $ t \sim 1/ |\alpha| $ (and similar periodic times) for $ \alpha = \beta $  hence in practice may only be effective in this region.  

Applying the Hillery-Zubairy criterion to the cross-Kerr case with operators (\ref{x1b2}) reveals no entanglement for all $ t $. This is due to the fact that one of the operators is Hermitian, for which the separability inequality can never be violated.  The DGCZ criterion can however detect entanglement \cite{crkerrduan}.  Choosing similar operators and following the same procedure as in Ref. \cite{duan} gives the criterion for separable states
\begin{align}
\delta(p_a + \tau t\alpha n_b)^2 + \delta(p_b + \tau' t\beta n_a)^2 \ge 
| \tau' t\beta \langle x_a \rangle | + |\tau t\alpha \langle x_b \rangle |  \label{duankerr}
\end{align}
where $ \tau $ and $ \tau' $ are parameters that can be optimized. From Fig. \ref{fig1}(b) we see that beyond short times $ t \ll 1/ \alpha $, entanglement cannot be detected using the DGCZ criterion. We attribute this to the strongly non-Gaussian nature of the states that are generated using the cross Kerr nonlinearity.  In deriving the DGCZ inequality, a postion-momentum type Heisenberg uncertainly relation is used, which is not necessarily the relevant relation for strongly non-Gaussian states.  In Fig. \ref{fig1}(d) we examine the dependence of characteristic times with the coherent state amplitude for our criterion and DGCZ.  For the DGCZ criterion we plot the time where entanglement is no longer is detected, while for our criterion we plot the time where the maximum violation is achieved.  We find that the scaling of our relation follows a power law as $ t \propto 1/ |\alpha| $, while the DGCZ criterion falls off at a faster rate than $ t \propto 1/ |\alpha|^2 $.  This shows that our criterion works in a considerably larger range of times, particularly as the amplitude of the coherent state is increased.

In summary, we have derived a correlation based entanglement witness for bipartite systems.  The criterion works with an arbitrary pair of observables on each of the subsystems.  We have compared the performance with other correlation based criteria, and find that in many cases it detects entanglement in a larger range of parameters than similar methods such as those of Hillery-Zubairy and Duan-Giedke-Cirac-Zoller.  The cost of this improvement is the necessity to evaluate a correlation involving the density matrix itself, which is related to the purity of the system.  We provide several methods of estimating this term, and find that it is possible under several circumstances to do so in a way that reduces the effectiveness of the criterion in a minimal way. We have investigated primarily the case (\ref{finalcri}) which is a special case of the family of inequalities (\ref{proof}) and (\ref{proof2}).  Due to the great flexibility of the general expression, there is much scope for further investigation of other criteria using different combinations of operators.  This may lead to more sensitive entanglement criteria, particularly for strongly non-Gaussian states which are less easy to detect using correlations based methods.

\begin{acknowledgments}
We thank Barry Sanders, Matteo Fadel, Shunlong Luo, Shaoming Fei, Fengxiao Sun, Yuxiang Zhang and Ming Li for discussions. Q.H. acknowledges the support of Ministry of Science and Technology of China (2016YFA0301302), the National Natural Science Foundation of China under Grants 11274025, 61475006, 11622428, and 61675007. T.B. acknowledges the support of the Shanghai Research Challenge Fund, New York University Global Seed Grants for Collaborative Research, National Natural Science Foundation of China grant 61571301, the Thousand Talents Program for Distinguished Young Scholars, and the  NSFC Research Fund for International Young Scientists. 
\end{acknowledgments}


\end{document}